\documentclass[twocolumn,pre,floatfix,superscriptaddress]{revtex4}
\usepackage{graphicx}
\usepackage{times}
\usepackage{color}
\usepackage{amsmath}
\usepackage{epstopdf}
\def\D{{\rm d}}

\begin{document}

\title{Reshaping nemato-elastic sheets}
\author{A.~P. Zakharov and L.~M. Pismen}
\affiliation{Department of Chemical Engineering, Technion -- Israel Institute of Technology, Haifa 32000, Israel}

\begin{abstract}
We consider three-dimensional reshaping of thin nemato-elastic sheets containing half-charged defects upon nematic-isotropic transition. Gaussian curvature, that can be evaluated analytically when the nematic texture is known, differs from zero in the entire domain and has a dipole or hexapole singularity, respectively, at defects of positive or negative sign. The latter kind of defects appears in not simply connected domains. Three-dimensional shapes dependent on boundary anchoring are obtained with the help of finite element computations. 

\end{abstract}
 
 \maketitle

Liquid crystal elastomers and glasses \cite{Warner}, made of cross-linked polymeric chains with embedded mesogenic structures, combine orientational properties of liquid crystals with shear strength of solids, and were envisaged by de Gennes as prototype artificial muscles \cite{degennes97}. Their specific feature is a strong coupling between the director orientation and mechanical deformations, which can be controlled by the various physical and chemical agents. When the material undergoes a phase transition from the isotropic to nematic state, it strongly elongates along the director and, accordingly, shrinks in the normal direction to preserve its volume; the opposite effect takes place as result of the reverse transition. Stresses arising due to these intrinsic deformations were investigated for flat sheets where they were shown to lead to phase separation of isotropic and nematic domains \cite{epj13} or formation of persistent defects that are not necessitated by topology \cite{pre15}.  

Internal stresses can be relaxed when three-dimensional (3D) deformations are allowed. The reshaping effect causes bending of flat thin sheets into curved shells. Aharoni \emph{et al} \cite{sharon} studied the various nematic textures that can be impressed in the material to induce a desired two-dimensional (2D) metric that would determine a 3D shape upon transition from nematic to  isotropic state (NIT). This opens the way to construct surfaces with nonzero Gaussian curvature out of flat sheets with no stretching \cite{Warner12}, or, more generally, to change the Gaussian curvature of shells. While  the Gaussian curvature is an intrinsic property that can be computed based a surface metric only \cite{Aris}, computing an actual 3D shape is a far more challenging task, which is still more complicated by constraints on nematic tissues imposed by boundary conditions or topology of closed manifolds that necessitate emergence of defects. A 3D shape can be computed analytically only for a symmetric texture,  and the only available example is deformation of a circle containing a defect with unit circulation into a cone or an anticone \cite{Warner12,pre14}. This case is exceptional in two respects. First, the natural nematic texture induced within a circle by either tangential or normal boundary anchoring contains two half-charged defects which have a lower energy than a single defect of unit charge. Second, the Gaussian curvature induced by a unit charge defect vanishes everywhere except the vicinity of the defect itself  \cite{sharon}, where it can be regularized either by local stretching \cite{Warner12} or by the depletion of nematic order \cite{pre14}. Away from the defect, where the Gaussian curvature vanishes, folding turns out to be necessary \cite{pre14} to accommodate an extended perimeter of the circle into a sphere with a shrinking radius, in a way similar to crumpling of isotropic non-extensible sheets \cite{witten}.

In a more difficult and realistic case of  half-charged defects, even planar textures can be computed analytically only in simplest configurations, and the lack of axial symmetry leaves little hope for analytical theory of 3D shapes. The Gaussian curvature differs then from zero also outside defect cores, which makes smooth 3D shapes possible. In this communication, we compute numerically the shapes generated upon NIT from naturally formed nematic textures. We first summarize our basic methods: analytical, starting with metric transformation induced by NIT and extending, in the case when the nematic texture is known, to computation of Gaussian curvature, and numerical, indispensable for visualizing actual 3D shapes. We consider next representative particular cases: a vicinity of isolated defects, where the induced structure can be well resolved; an elliptic sheet containing two half-charged defects; and a sheet with two holes where negatively charged defects naturally appear.  

We consider a thin sheet with preferred parallel orientation of the director on both outer surfaces. The material is described then by the two-dimensional 2D nematic order parameter, presented as a traceless symmetric tensor dependent on 2D coordinates:
\begin{equation}
 Q^\alpha{}_\beta = S\left(n^\alpha n_\beta -\mbox{$\frac 12$}\delta^\alpha_\beta\right),
 \label{eq:qn}
\end{equation}
where $n_{\alpha}$ are vector components of the director, $\delta^\alpha_\beta$ is the Kronecker delta, and $S$ is the scalar order parameter equal to unity in a perfectly aligned nematic state. 
Under the assumption of negligible elastic deformations, an infinitesimal interval is transformed as a result of a change $\widetilde{Q}^\alpha{}_\beta$ of the nematic state as 
\begin{equation}
\D\xi^\alpha=T^\alpha{}_\beta \D x^\beta, \quad
T^\alpha{}_\beta = N^{-1/2} \left(\delta^\alpha_\beta
 + a \widetilde{Q}^\alpha{}_\beta \right),  
\label{eq:xxi}
\end{equation}
where $a$ is a metric factor measuring the relative extension and contraction, and $N$ is a normalization factor; summation over paired indices is presumed. The original metric tensor $\mathbf{g}^0$ is transformed thereby to $\mathbf{g} =\mathbf{Tg}^0\mathbf{T}$; the transformation can be made area-preserving \cite{pre14} by choosing $N$ to ensure det(\textbf{T}) $=1$. In the case of transition from nematic ($S=1$) to isotropic ($S=0$) state (NIT), the explicit metric expression is 
\begin{equation}
g_{\alpha\beta}=(1-a^2)^{-1} \left[(1+a^2)g^0_{\alpha\beta}
 - 2a Q_{\alpha\beta} \right).  
\label{eq:ggam}
\end{equation}
If the coordinate axes $x^\alpha$, $\xi^\alpha$ are oriented, respectively, along and normal to the director, Eq.~\eqref{eq:xxi} simplifies to  
\begin{equation}
\D\xi^1=\frac{1}{\ell}\D x^1, \quad \D\xi^2=\ell \,\D x^2, 
 \quad \ell=\sqrt{\frac{1+a}{1-a}},  
\label{eq:x12}
\end{equation}
and the transformed metric is diagonal with the elements $\ell^{-2}, \,\ell^2$.

The transformed metric can be used to compute \emph{intrinsic} properties of bent sheets or shells but not their extrinsic properties defining their position in space. Among the former, the most important one is Gaussian curvature equal to the only non-zero component $K=R_{1212}$ of the Riemann curvature tensor \cite{Aris}. 
It is defined by standard formulas dependent on the first and second derivatives of the metric tensor. 

Computing actual 3D shapes which, unlike Gaussian curvature, are not determined by intrinsic characteristics of a surface alone, is a much more difficult task that can be only accomplished numerically. We carried out computations by triangulating the surface and defining the nematic order parameter at the grid nodes. 
The nematic director on triangular tiles was defined as the average of the three vertices. Normal vectors $\mathbf{m}_i$ were computed for each tile, and the normal vectors at nodes $\mathbf{m}_j$ were defined as the average of the normals at the surrounding tiles. If the nematic field was known analytically, as in some following examples, \ref{S:ell}, the nematic director was defined at the centres of triangular tiles. 

The triangles were addressed at random following a Monte Carlo routine, and NIT was imitated by changing the length of their sides (edges) according to Eq.~\eqref{eq:xxi} and further rescaling to eliminate numerical errors in maintaining area conservation. The procedure continued, typically, for 20 Monte Carlo cycles to minimize the energy functional defined as 
\begin{equation}
E = \sum_\mathrm{edges} (l_i - l_i^0)^2 
+ \kappa \sum_\mathrm{nodes} \sum_j (\mathbf{m}_i\cdot \mathbf{m}_{ij}). 
 \label{eq:MC}
\end{equation}
Here, the first term accounts for deviation of the length $l_i$ of an $i$th edge from the value $l_i^0$ defined by Eq.~\eqref{eq:xxi}. In the second term, $\mathbf{m}_{ij}$ denotes the normal to the a $j$th tile of those surrounding an $i$th node. The sum over all neighbors measures the deviation from their average orientation $\mathbf{m}_{i}$, which accounts for the curvature of the bent shell. Since bending rigidity is much weaker that in-shell shear modulus, the ratio $\kappa$ of bending rigidity to in-shell elasticity should be small. In order to speed up numerical algorithm at early stages, it was gradually decreased from $10^{-4}$ to $5 \times 10^{-6}$, which allowed us to arrive as close as possible to the required degree of local extension or shrinkage.     

When the nematic texture was not known analytically, we computed it in one-constant approximation by minimizing the nematic energy $\int (\nabla \mathbf{n})^2 d\Sigma$ over a studied domain with appropriate anchoring conditions at boundaries. The derivatives are approximated by differences between the directors at neighboring nodes. Although the director is more naturally defined on tiles, redefining it on nodes improves numerical accuracy due to a larger number of neighbors (six rather than three).  We resolved the defect core structure only in the case of isolated defects leaving otherwise the scalar order parameter $S$ equal to unity, but triangulation density was increased in the vicinity of defects. An amended procedure is applicable to computation of textures on curved surfaces. In this case, covariant derivatives are computed by virtually rotating neighboring tiles into a common plane before comparing the directors, and an additional energy penalty is introduced according to the surface curvature along the director.

\begin{figure}[b]
\centering 
\includegraphics[width=.23\textwidth]{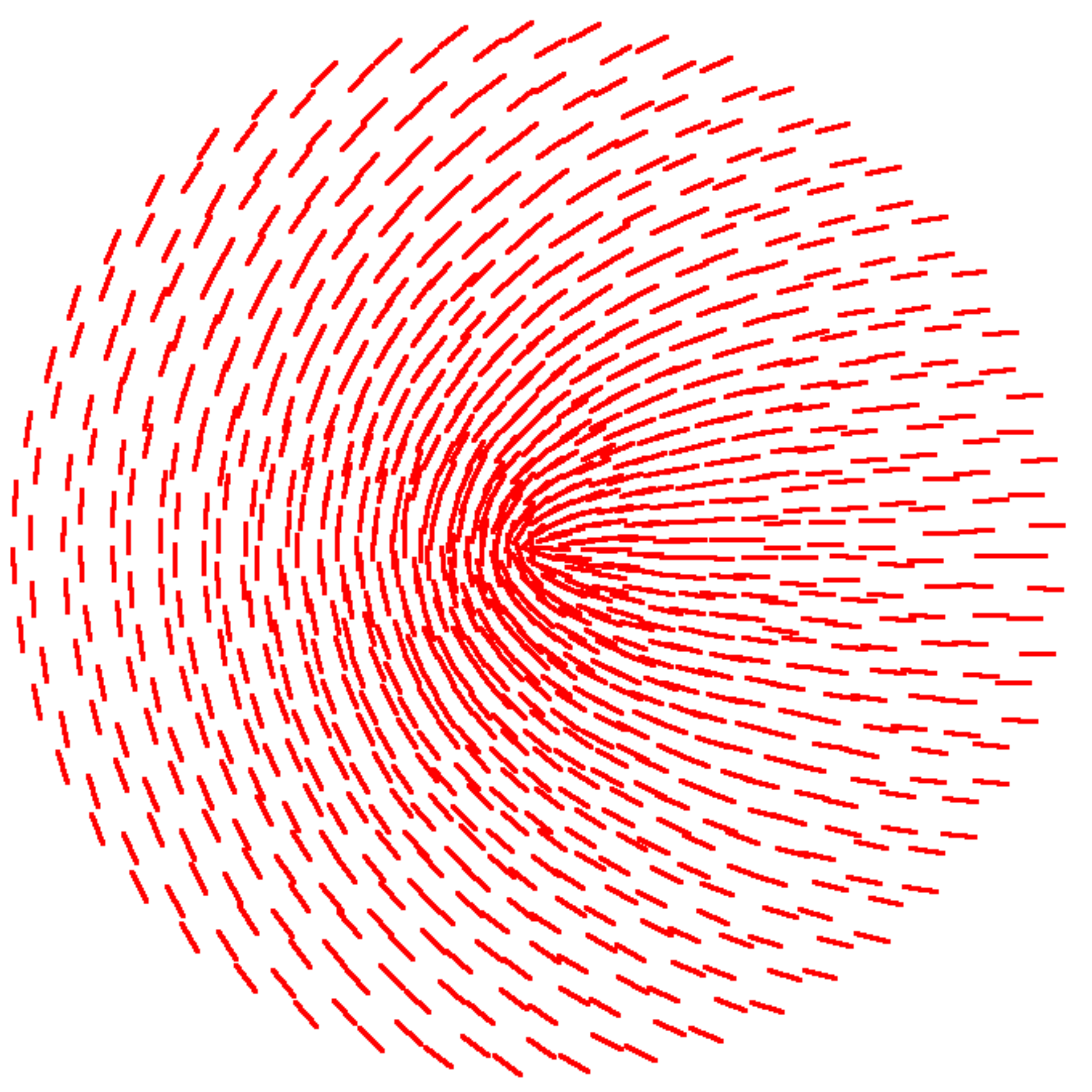}
\includegraphics[width=.23\textwidth]{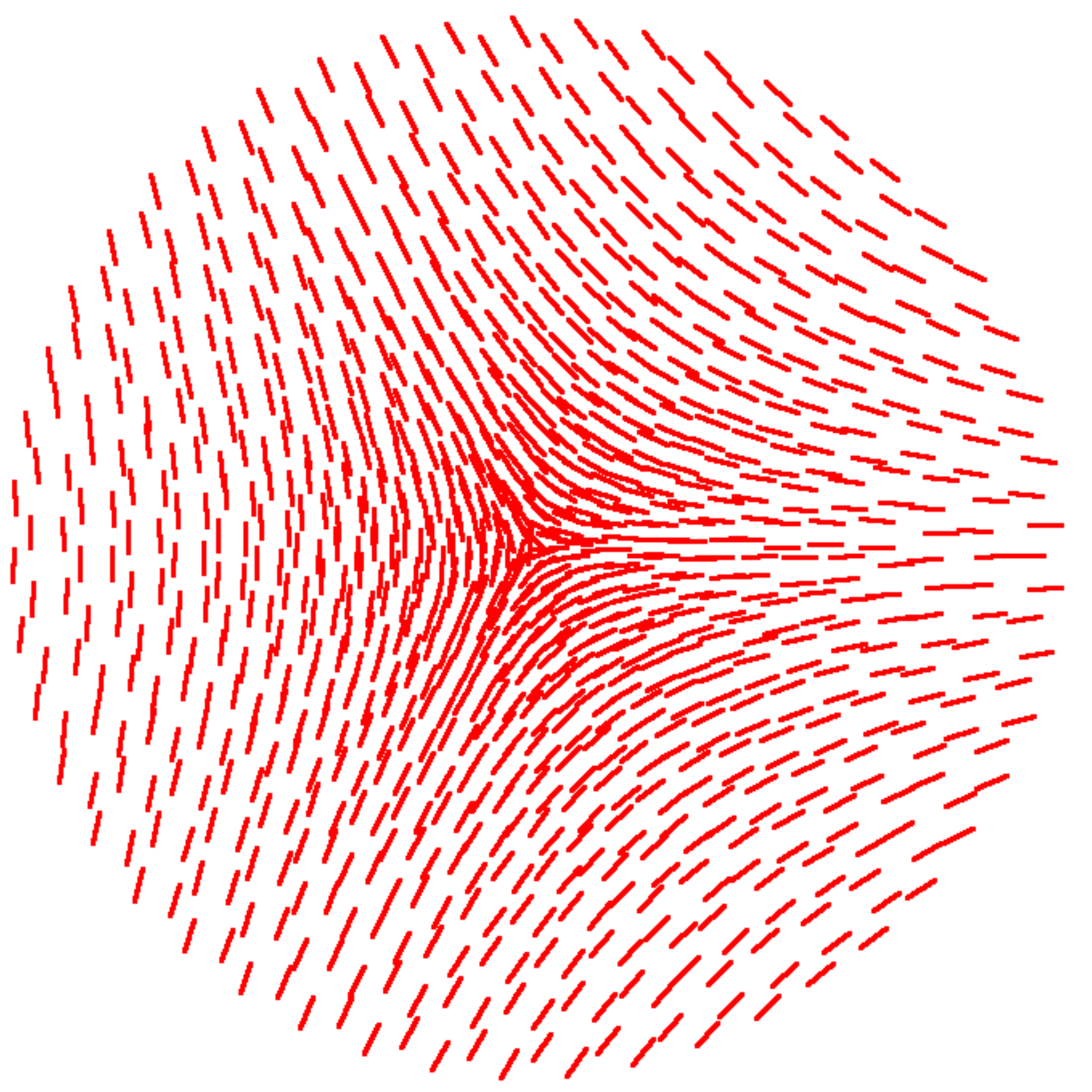}
\includegraphics[width=.24\textwidth]{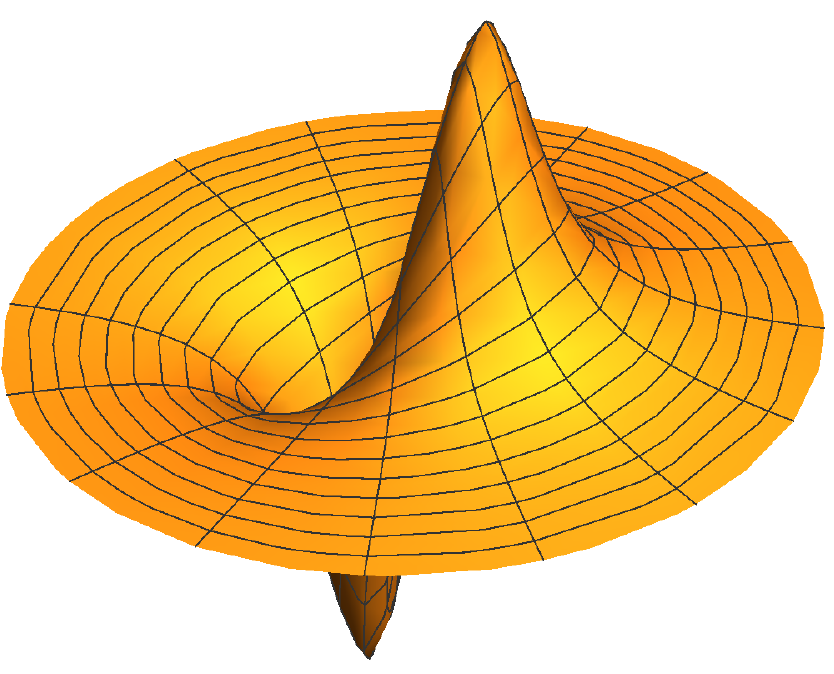}
\includegraphics[width=.23\textwidth]{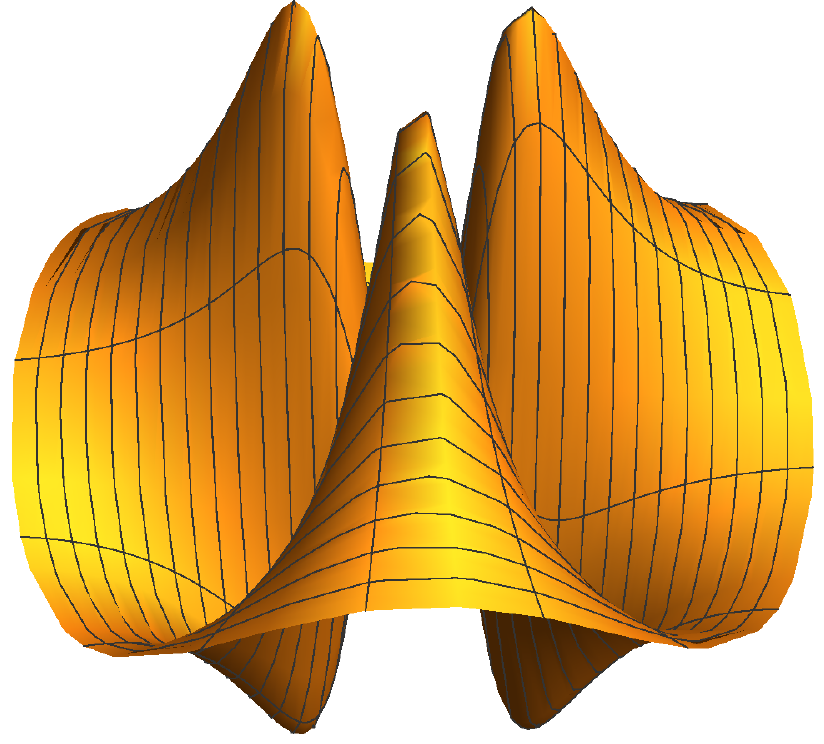}
\includegraphics[width=.24\textwidth]{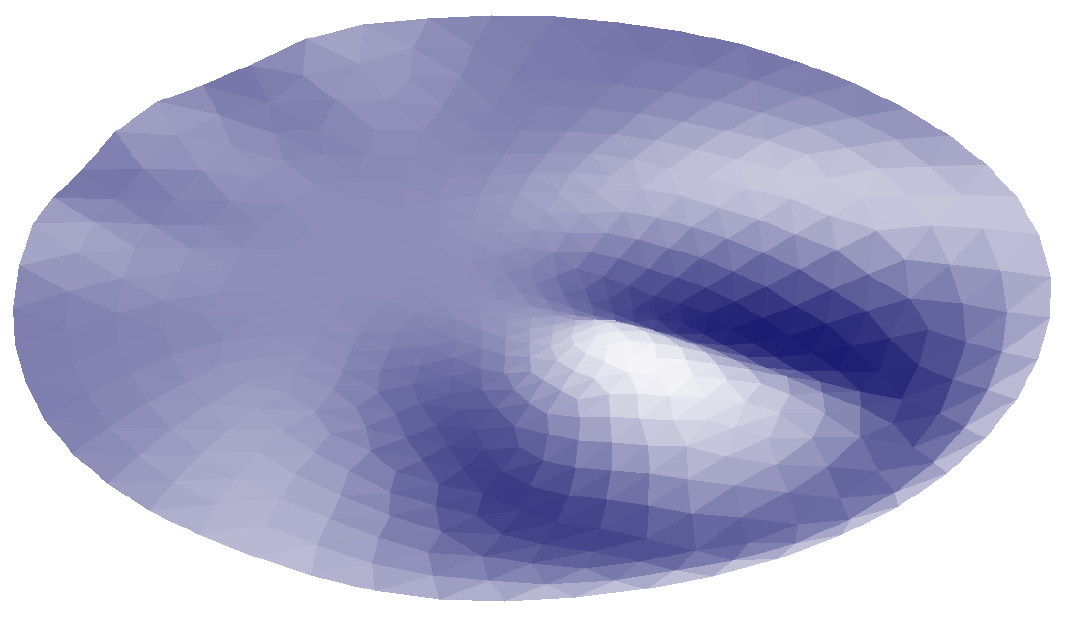}
\includegraphics[width=.23\textwidth]{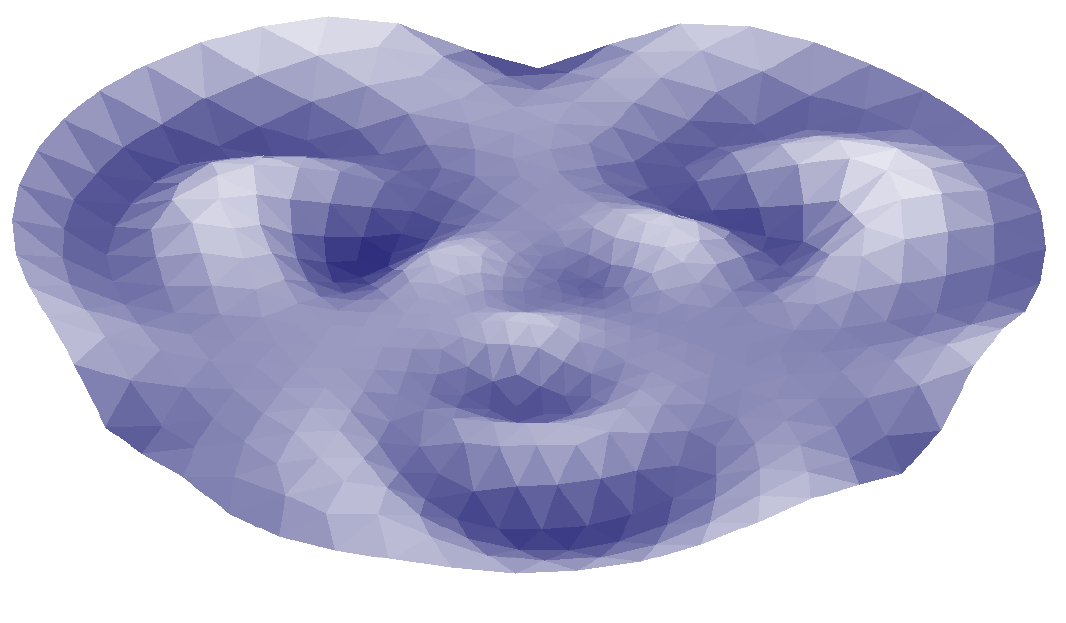}
\caption{The vicinity of isolated positive (left) and negative (right) half-charged defects. Above: the director field; middle row: the regularized Gaussian curvature; below: 3D shape.}
\label{def1}
\end{figure}
Strong reshaping effects in the vicinity of defects are of special interest, and can be studied by considering isolated defects, far removed from other defects and boundaries. The texture in the vicinity of positive and negative half-charged defects has different symmetry (Fig.~\ref{def1}), even though the nematic order parameter is expressed in a similar way. In a vicinity of defects of the charge $\pm 1/2$ (but outside their cores) the tensor order parameter is 
\begin{equation}
\mathbf{Q}= \frac{1}{2} \left(\begin{array}{cc}\cos \phi & \pm\sin \phi \\ 
 \pm\sin \phi & -\cos \phi \end{array} \right), \label{eq:Q12}\end{equation}
where $\phi$ is the polar angle. Starting from the Cartesian metric $\mathbf{g}^0$, we obtain using Eq.~\eqref{eq:ggam} the transformed metric 
\begin{equation}
\mathbf{g} = \frac{1}{1-a^2} \left(\begin{array}{cc}1-2a\cos \phi+a^2 
 & \mp 2a\sin \phi \\ 
 \mp 2a\sin \phi & 1+2a\cos \phi+a^2 \end{array} \right).
 \label{eq:g12}
\end{equation}
The Gaussian curvature $K_\pm$ of a bent shell formed upon NIT is derived from this metric:
\begin{equation}
K_+ = \frac{a \cos \phi}{r^2(1-a^2)}, \qquad
K_- = -\frac{3a \cos 3\phi}{r^2(1-a^2)}.
 \label{eq:g12}
\end{equation}
These simple expressions are in accordance with the symmetry of the bent shell around the defect. The curvature diverges at the defect location ($r \to 0$) leading to a dipole or hexapole singularity, respectively, for defects of positive or negative sign.  For the positive charge, the curvature is positive near the defect's ``comet tail'' at $\phi=0$, while a saddle geometry with negative Gaussian curvature is indicated on the opposite side. For the negative charge, there is a threefold symmetry, matching the symmetry of the nematic field (Fig.~\ref{def1}). 

The divergence can be regularized by taking into account vanishing $S$ at the defect core. For defects of either sign, the radial dependence $S(r)$ is identical to that of the amplitude of a complex scalar field near a defect of \emph{unit} charge \cite{pre13}, and satisfies the equation
\begin{equation}
 S''(r) +r^{-1}  S'(r) + S \left(1 - r^{-2} -S^2 \right) =0,
\label{eq:evolps}
\end{equation}
where the radial coordinate is scaled by the healing length. The regularized Gaussian curvature within a circle encompassing the defect core is shown in Fig.~\ref{def1}.

\begin{figure}[b]
\includegraphics[width=.23\textwidth]{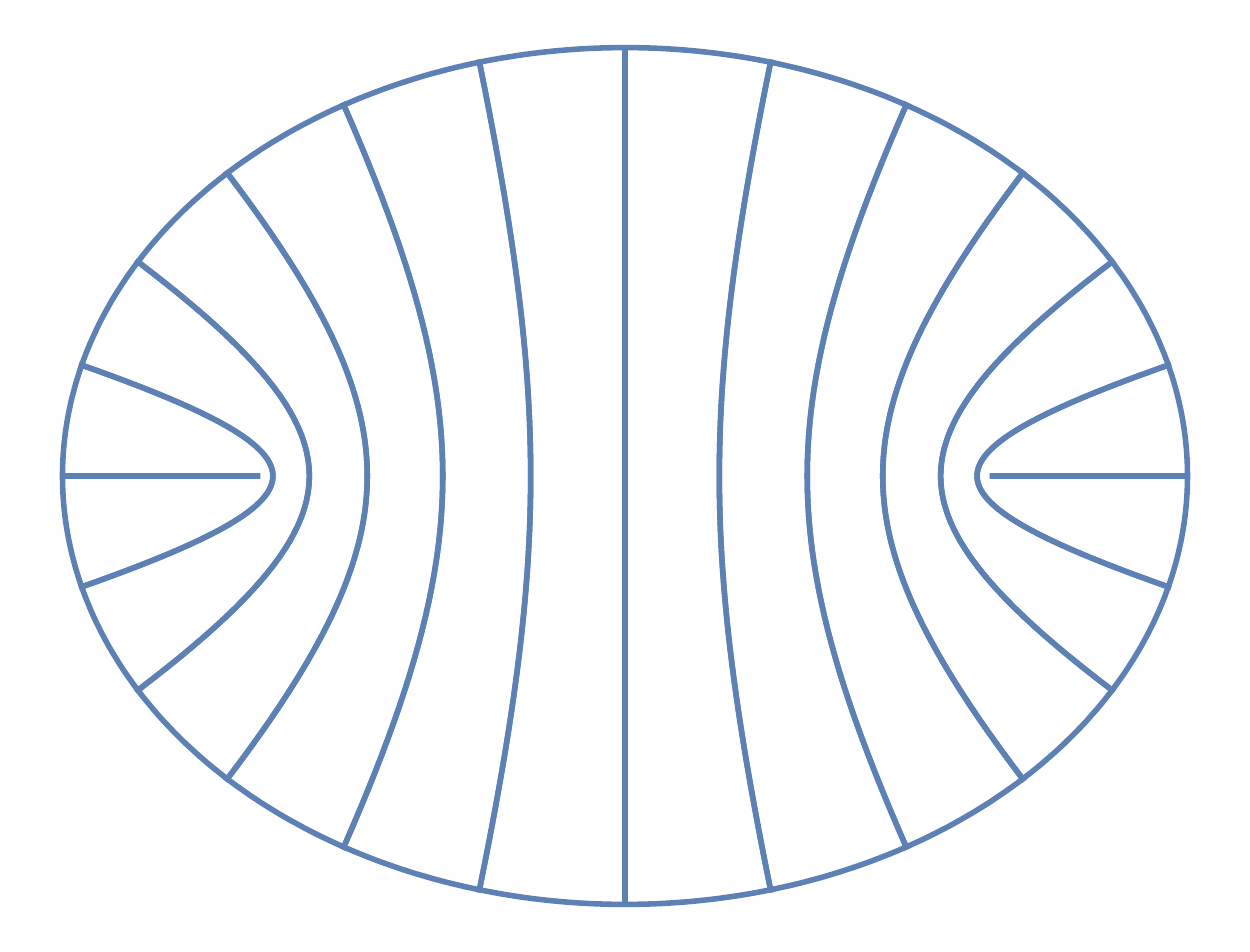}
\includegraphics[width=.23\textwidth]{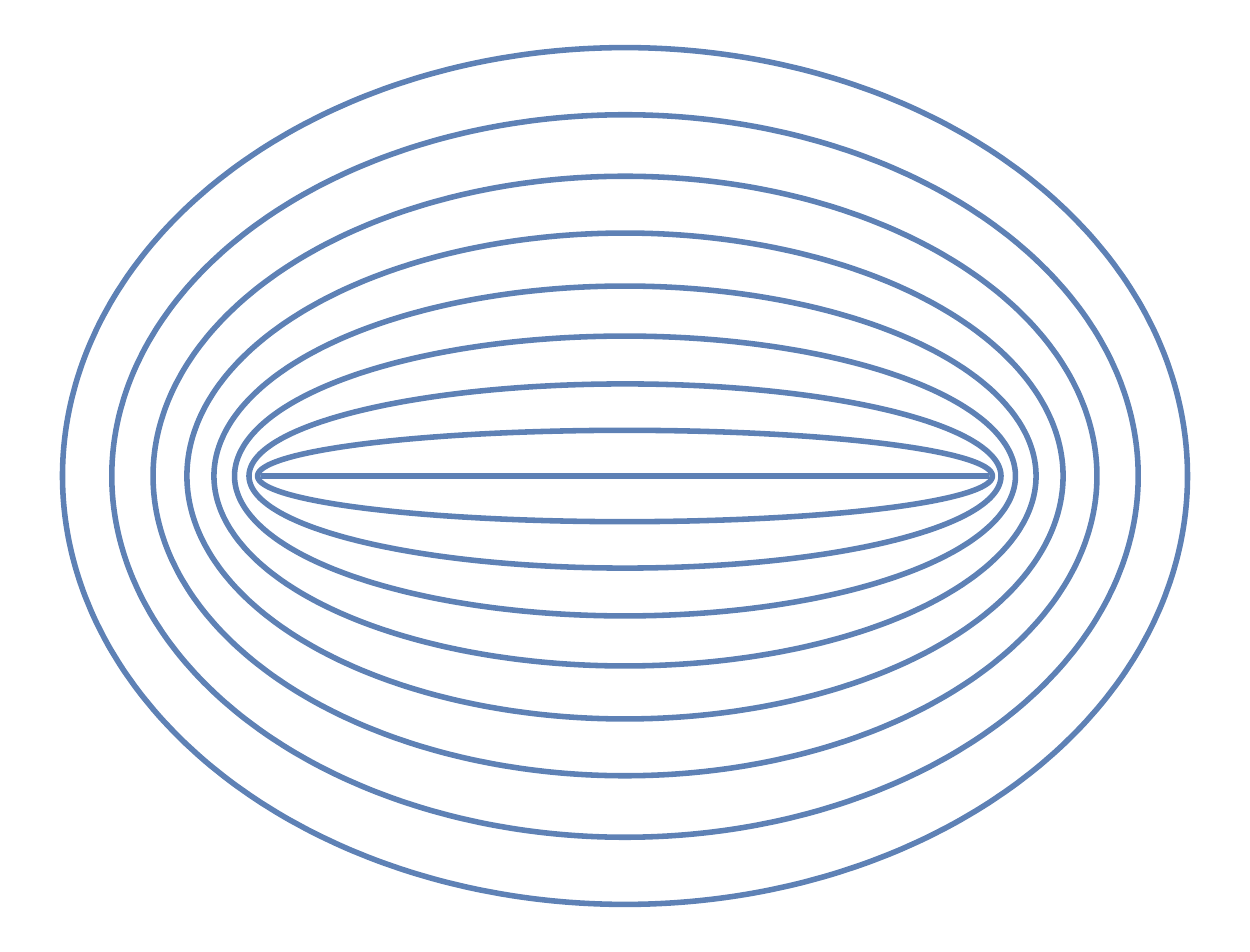}
\includegraphics[width=.37\textwidth]{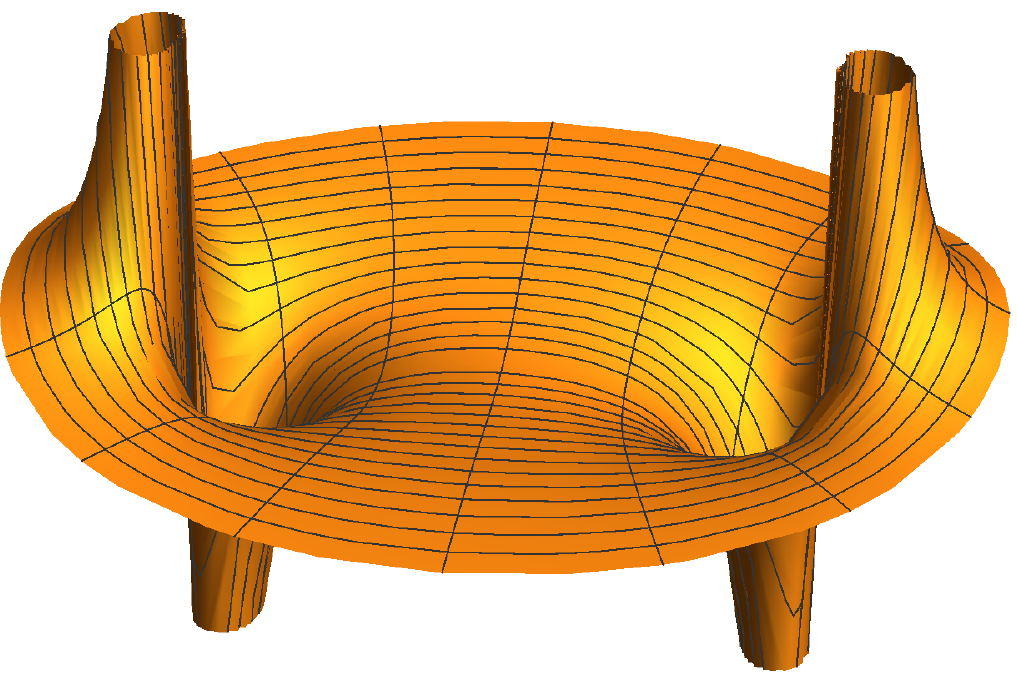}\\
\caption{Above: director orientation for normal and tangential boundary anchoring. Below: Gaussian curvature. }
\label{f:ellipse}
\end{figure}
\begin{figure}[t]
\includegraphics[width=.45\textwidth]{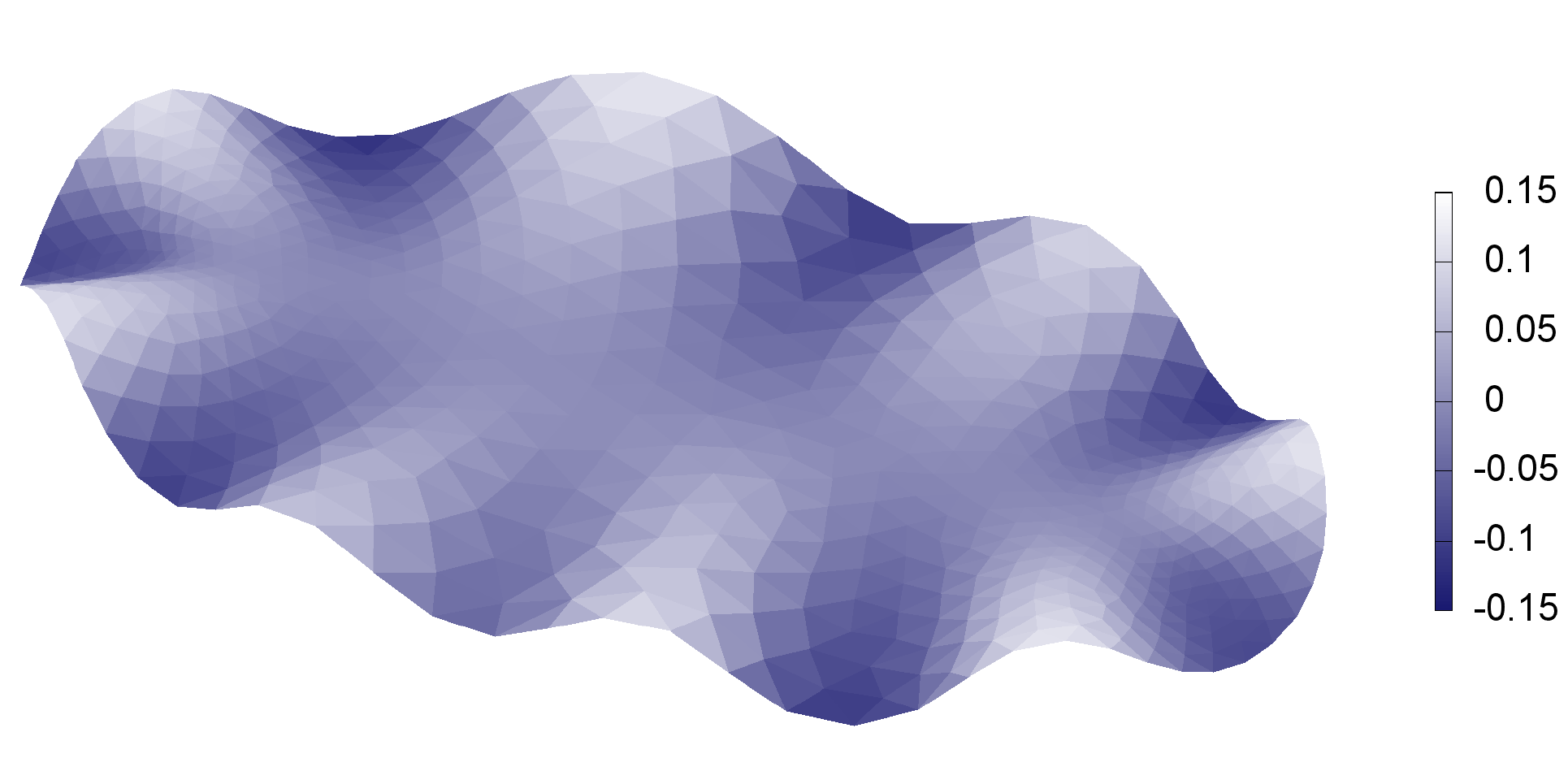}\\
\includegraphics[width=.45\textwidth]{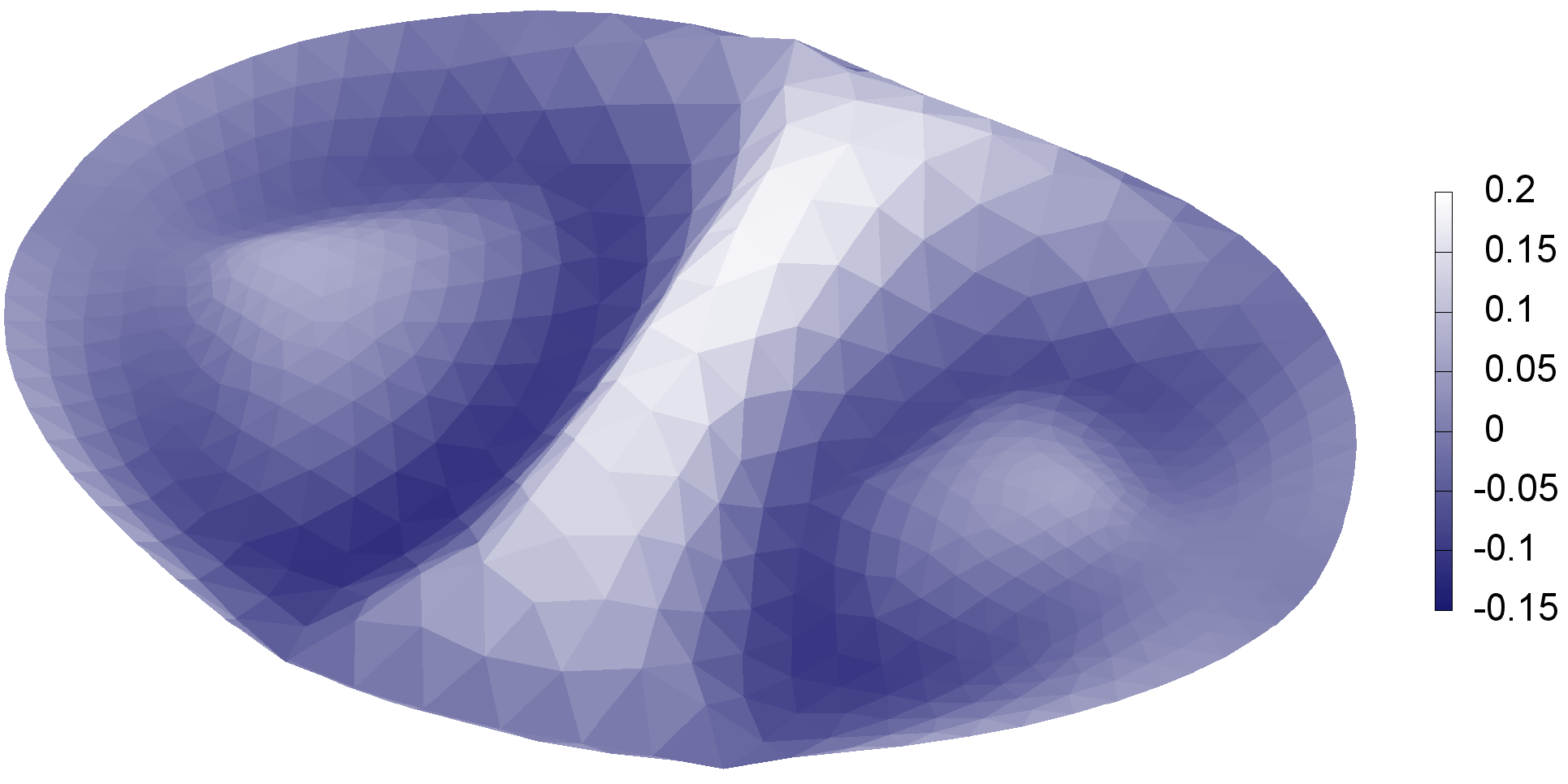}\\
\includegraphics[width=.42\textwidth]{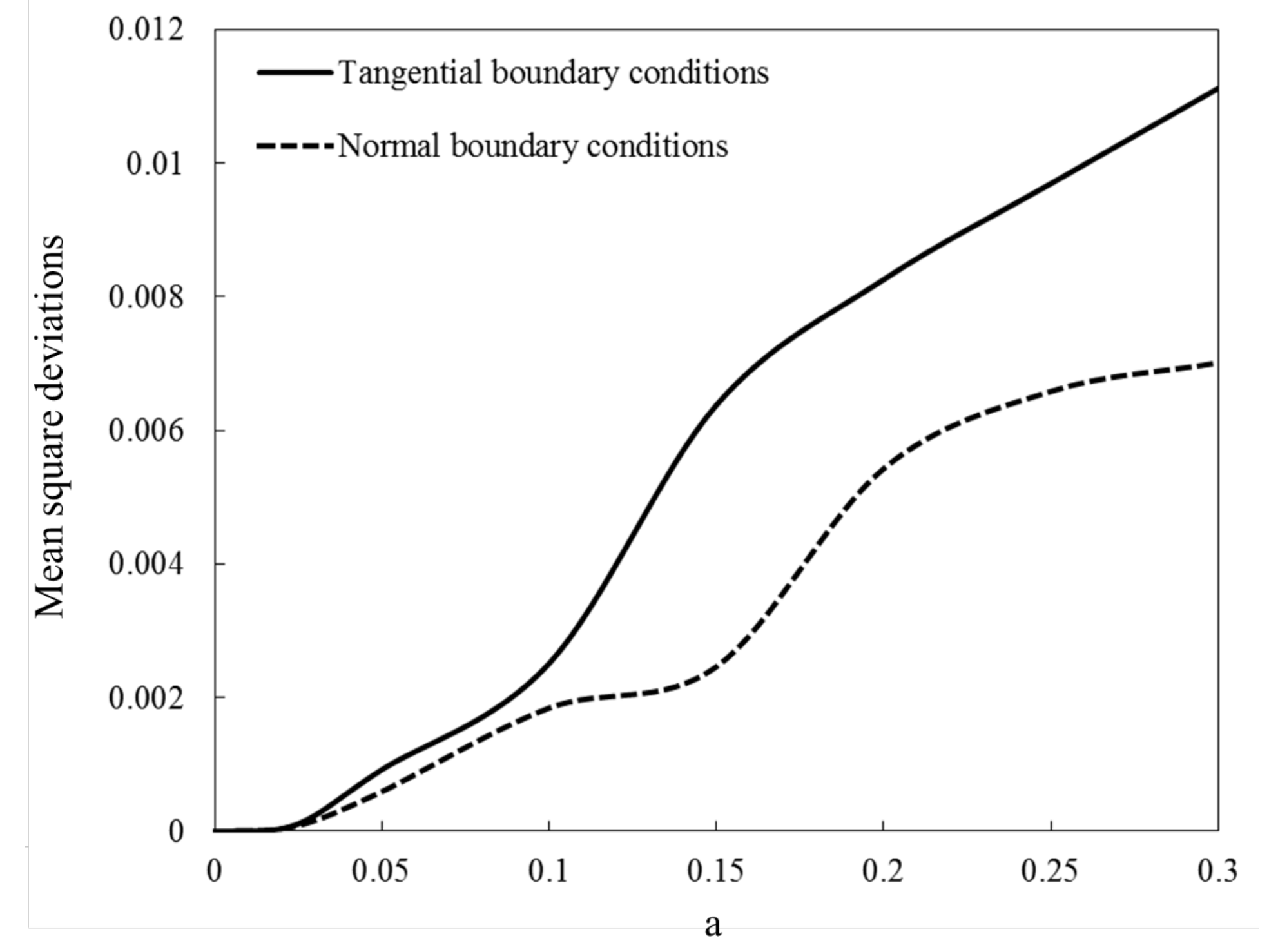}
\caption{Deformation of an ellipse. Above: 3D shapes for normal anchoring and tangential anchoring ($a=0.1$). Below: the dependence of the mean square deviation of the surface from the original position as a function of the metric factor $a$.}
\label{f:ellipsed}
\end{figure}
In only two known examples, nematic texture containing half-charged defects can be obtained analytically in the one-constant approximation by means of conformal transformation \cite{nelson}. The simplest case is an ellipse where the director is aligned along coordinate lines $u=$ const (if it is anchored normal to the boundary) or $v=$ const (if the anchoring is tangential) of the elliptic coordinate system, which are identical to the isolines of the real and imaginary part of the analytic complex function $w=\sin z$ (Fig.~\ref{f:ellipse}). Upon NIT, the metric of the resulting curved sheet is readily obtained by modifying the elliptic metric according to Eq.~\eqref{eq:x12}:      
\begin{equation}
\mathbf{g} =g\, \mathrm{diag}\left(\ell^{-2}, \ell^2 \right),
\quad  g= \frac 12 (\cosh 2v -\cos 2u),
 \label{eq:ellipseg}
\end{equation}
where the half-focal distance is normalized to unity. This expression is equally applicable to textures with the director along the coordinate lines $u=$ const or  $v=$ const; one needs only to interchange the metric factors corresponding to the two variables. In both cases, the Gaussian curvature is computed as 
\begin{align}
K  &=\pm\frac 1g \left[\frac {\partial} {\partial u }\left(\frac 1g \frac{\partial g} {\partial u }\right) + \frac {\partial} {\partial v }\left(\frac 1g \frac{\partial g} {\partial v }\right)\right] \notag \\
 &= \pm\frac {16a(\cosh 2v \cos 2u-1)}{(1-a^2)(\cosh 2v -\cos 2u)^2}.
 \label{eq:ellipseH}
\end{align}
This expression has a dipolar singularity at both foci (see Fig.~\ref{f:ellipse}). Near the foci at $x=\pm 1, \, y=0$, one can set $u \approx \pm \sqrt{2r} \sin(\phi/2), \; v \approx \pm \sqrt{2r} \cos(\phi/2)$ and recover the first formula in \eqref{eq:g12} in the leading order of expansion in the distance $r \ll 1$ from either focus. The difference is in the position of ``comet tails'' of both defects, which are directed toward the centre of the ellipse when the anchoring is tangential, and outwards when the anchoring is normal. The positive and negative curvature regions are situated accordingly, so that the Gaussian curvature plot in Fig.~\ref{f:ellipse} corresponds to the normal anchoring and has to be flipped over when the anchoring is tangential. The shapes for both cases shown in Fig.~\ref{f:ellipsed} are quite dissimilar. The lower panel in Fig.~\ref{f:ellipsed} shows the nonlinear dependence of the mean square deviation of the surface from the original position as a function of the metric factor $a$.

\begin{widetext}
\begin{figure}[b]
\centering 
\begin{tabular}{cc} 
\includegraphics[width=.5\textwidth]{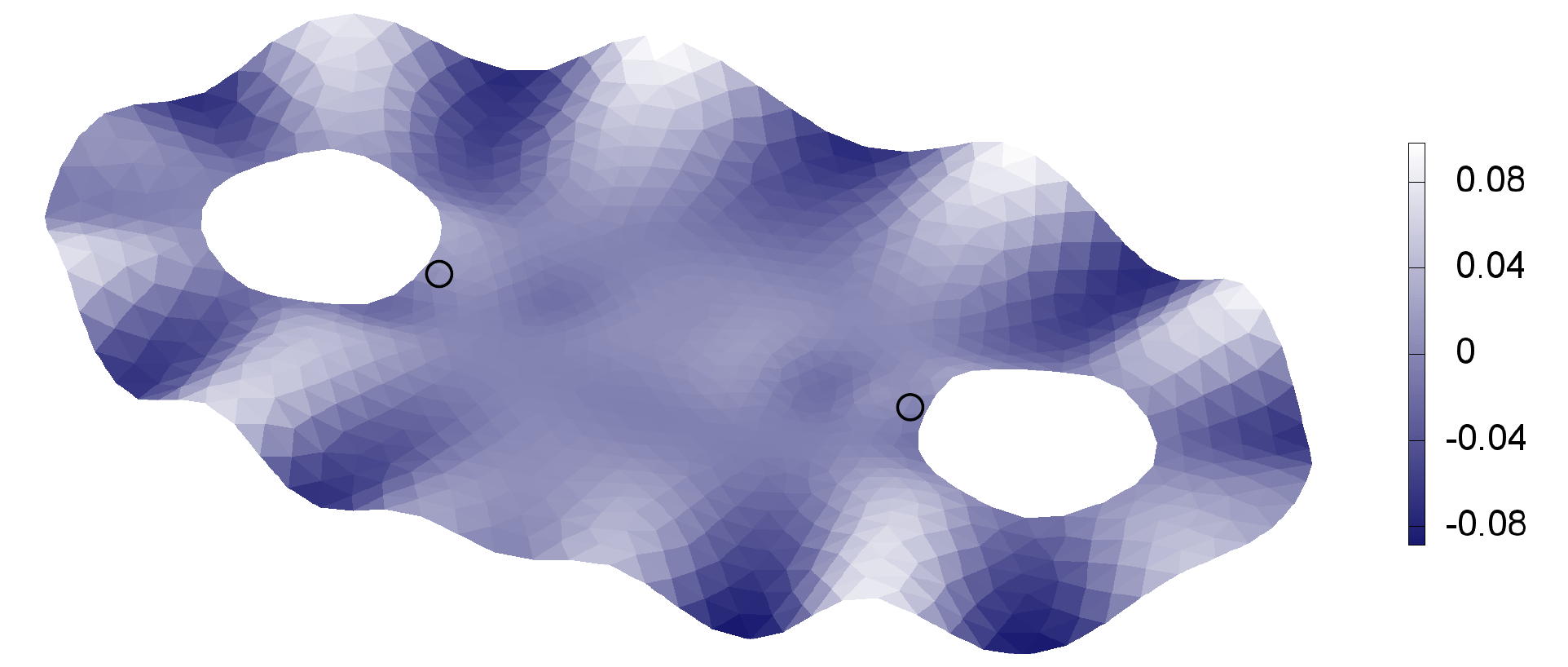} &\includegraphics[width=.5\textwidth]{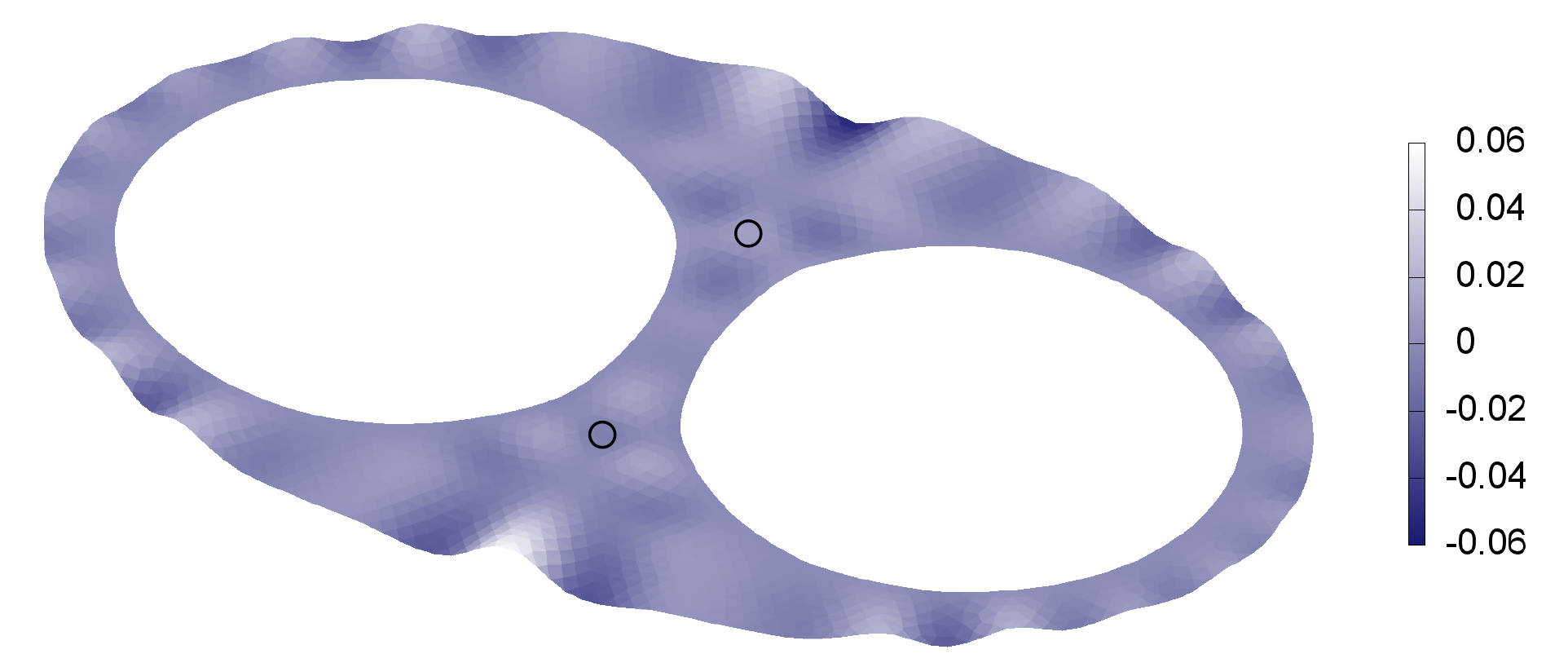}\\
\includegraphics[width=.5\textwidth]{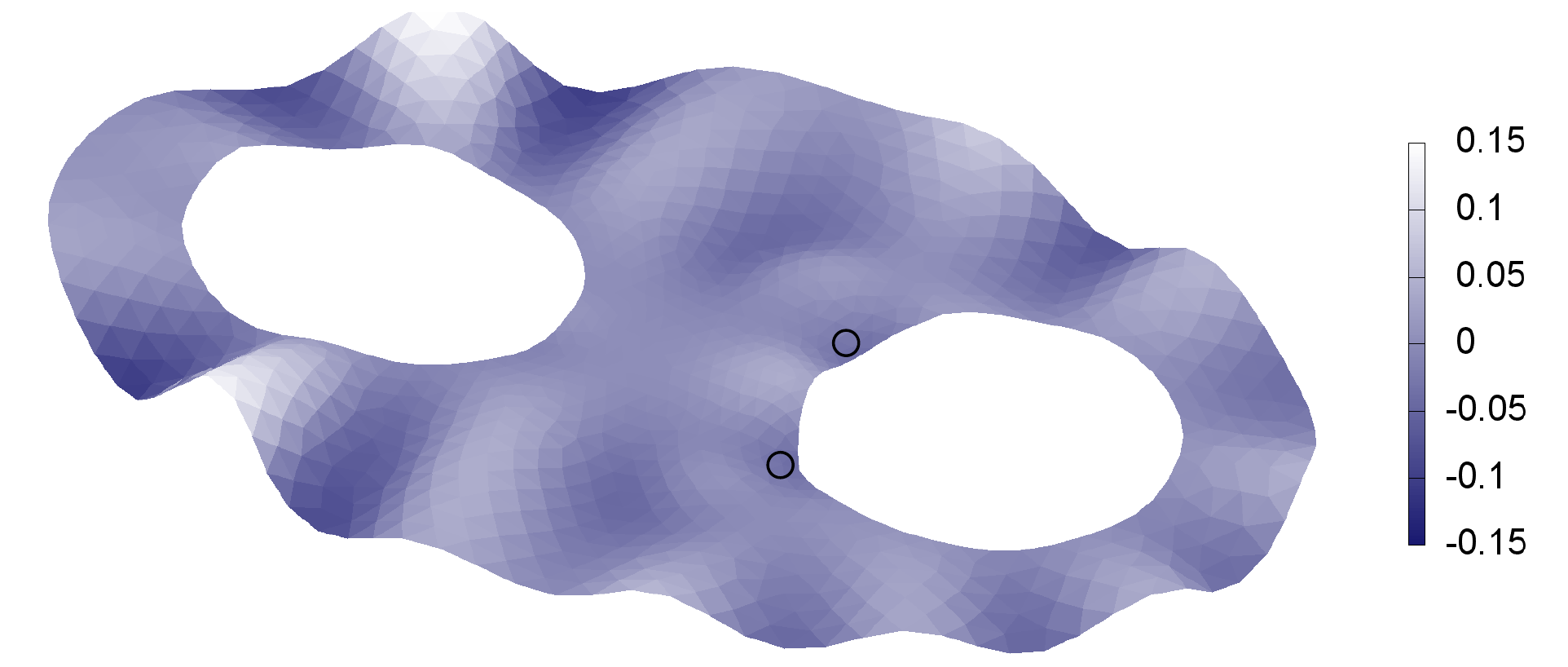}&
\includegraphics[width=.5\textwidth]{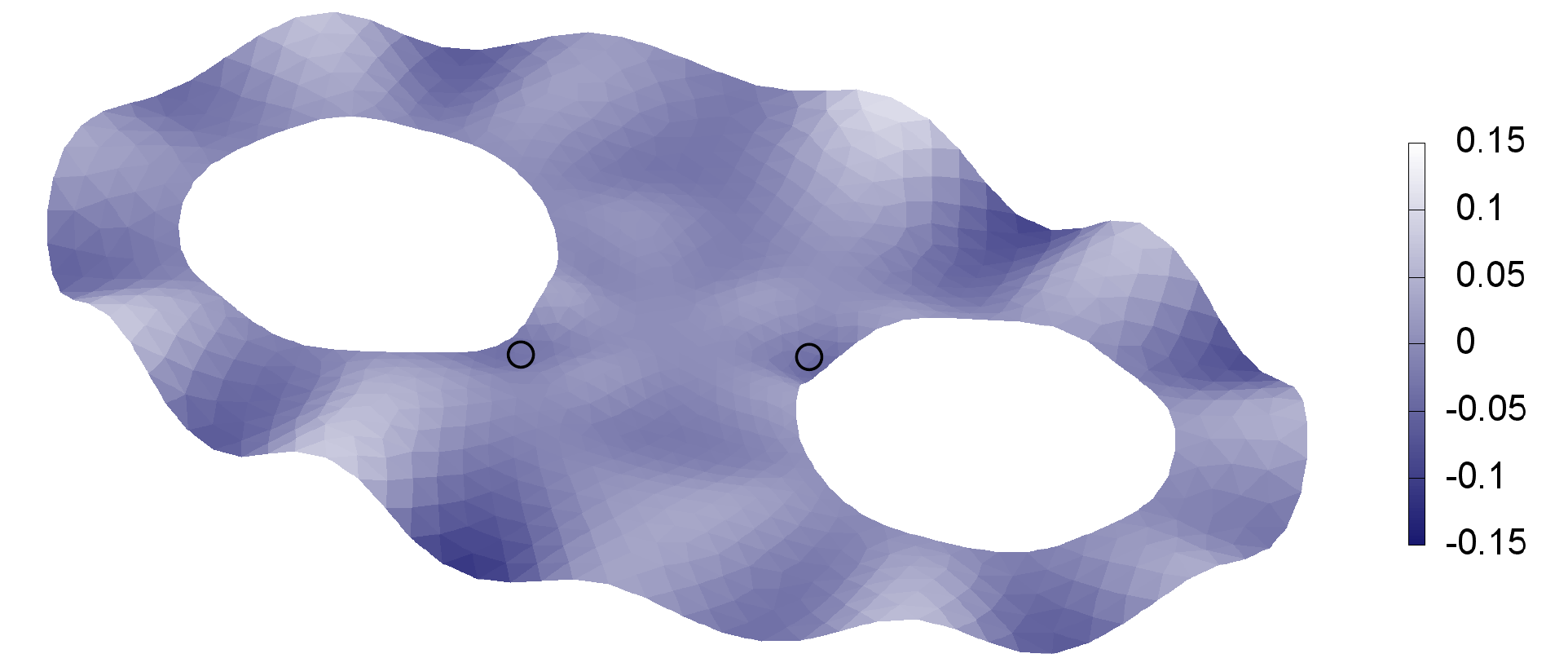}
\end{tabular}
\caption{Deformation of domains with two holes ($a=0.1$). Small circles mark positions of defects with the charge $-1/2$}
\label{f:2holes}
\end{figure}
\end{widetext}

Negatively charged defects appear when domains are not simply connected.  Inserting a single hole with the same anchoring as on the outer boundary leads to a lowest-energy texture with no defects. When a second hole is added, two defects with the charge $-1/2$ must appear. The position of defects depends on the size and position of holes. Some typical examples of elliptic domains with symmetrically placed two holes, as they deform following NIT, are shown in Fig.~\ref{f:2holes}. In all cases, the anchoring at all boundaries is normal. 

In the upper left picture with relatively small holes, the defects are situated symmetrically on the line connecting the centres of both holes. In the upper right picture, where the holes are large, the defects are placed on the symmetry axis separating the holes. The arrangement of defects in the two lower pictures with holes of intermediate size is less intuitive. In the left picture, the defects are situated asymmetrically near one hole, while in the right picture, they are placed diagonally near alternative holes keeping a larger mutual separation. The latter arrangement has a lower energy, so that its counterpart is \emph{metastable}. The corrugated shapes obtained following NIT, are, naturally, different in all cases.

Reshaping of nemato-elastic sheets or shells opens ways of creating a variety of forms that can be manipulated by boundary anchoring, positioning of defects, and topological changes. Besides static reshaping, the forms can be actuated dynamically, thereby creating crawling and swimming micro robots that will be the subject of a further study.

\emph{Acknowledgment}. This research is supported by Israel Science Foundation  (grant 669/14). We thank Michael K\"opt for stimulating discussions.


\end{document}